# COMPLEX-VALUED PHASE SYNCHRONY REVEALS DIRECTIONAL COUPLING IN FMRI AND TRACKS MEDICATION EFFECTS

*Sir-Lord Wiafe[1], Najme Soleimani[1], Masoud Seraji[1], Bradley Baker[1], Robyn Miller[1], Ashkan Faghiri[1], Vince D. Calhoun[1]*

[1]Tri-Institutional Center for Translational Research in Neuroimaging and Data Science (TReNDS), Georgia State University, Georgia Institute of Technology, and Emory University, Atlanta, GA 30303, USA.

## ABSTRACT

Understanding interactions in complex systems requires capturing the relative timing of coupling, not only its strength. Phase synchronization captures this timing, yet most methods either reduce the phase to its cosine or collapse it into scalar indices such as the phase-locking value, discarding relative timing. We propose a complex-valued phase synchrony (CVPS) framework that estimates phase with an adaptive Gabor wavelet and preserves both cosine and sine components. Simulations confirm that CVPS recovers true phase offsets and tracks non-stationary dynamics more faithfully than Hilbert-based methods. Because antipsychotics are known to modulate the timing of cortical interactions, they provide a rigorous context to evaluate whether CVPS can capture such pharmacological effects. CVPS further reveals cortical neuro-hemodynamic drivers, with occipital-to-parietal and prefrontal-to-striatal lead–lag flows consistent with known receptor targets, confirming its ability to capture pharmacological timing. CVPS, therefore, offers a robust, generalizable framework for detecting relative timing in complex systems such as the brain.

## 1. INTRODUCTION

Estimating how distributed systems exchange information requires not only measuring whether they are coupled, but also when one signal leads or lags another [1]. In neuroscience, communication is often inferred from functional connectivity [2]; yet, most metrics collapse timing asymmetries into symmetric correlations, erasing the relative timing that many systems use to coordinate activity [3]. Phase synchronization (PS) is an attractive alternative because instantaneous phase differences directly encode temporal alignment [4, 5]. Most pipelines, however, either retain only the cosine of the phase difference, cosine of relative phase (CRP) [5], reducing relative timing coupling to a symmetric time-resolved coupling strength [6, 7] or collapse the full complex angle into a scalar synchrony-strength measure such as the phase-locking value, mean phase coherence, or weighted phase lag index [5]. In both cases, the signed lag information is lost. Also, phase is commonly estimated using a Hilbert transform after fixed band-pass filtering, which smears non-stationary signals and underperforms in noisy, finite-length recordings [8].

We address these issues by proposing a complex-valued phase synchrony (CVPS) metric that estimates instantaneous phase using a complex Gabor wavelet and preserves both real and imaginary parts of synchrony between signals. This approach provides sharper joint time–frequency localization and greater robustness to non-stationary noise compared to Hilbert filtering. Each edge is encoded as a two-component vector $[\cos(\theta), \sin(\theta)]$, preserving both coupling strength and signed lag.

As a rigorous test case, we apply CVPS to functional magnetic resonance imaging (fMRI) recordings of schizophrenia and antipsychotic medication effects. Antipsychotics are hypothesized to retime cortical processes rather than the strength of coupling [9], making them an ideal validation scenario for a method designed to capture phase lead–lag structure. Importantly, no existing dynamic functional connectivity (dFC) pipeline has demonstrated a dose–response effect in a single cross-sectional scan: conventional approaches consistently report chlorpromazine-equivalent (CPZ) dose as non-significant [10, 11], forcing longitudinal protocols that require strict dose stabilization and repeated imaging [9, 12]. By contrast, we show that CVPS uncovers a reproducible brain state whose dwell time scales with CPZ dose after adjusting for symptoms and confounds. We also show that this effect is invisible to cosine-only PS and correlation-based dFC across several window sizes, highlighting the relevance of preserving the full complex phase as a timing-sensitive biomarker inaccessible to traditional pipelines.

To validate that the CPZ-linked state captures a genuine circuit modulation, we compute circular-distance-weighted driver scores that reveal network leaders and followers while avoiding the orthogonality constraints. This analysis is motivated by pharmacological evidence that antipsychotics act on parietal and early visual cortices, where 5-HT2A and $D_2$ receptor densities are highest [13, 14], as well as on cortico-striato-thalamo loops implicated in dopaminergic regulation [15]. Because fMRI phase captures the combined timing of neural and hemodynamic processes, CVPS should be understood as indexing their integrated lead–lag dynamics rather than purely neural conduction delays. Within this framework, we expect CVPS to capture pharmacodynamic retiming by identifying driver profiles that reveal visual-to-parietal lead flow and prefrontal dominance over subcortical receivers, consistent with known neuroreceptor targets.

Beyond psychiatry, these results demonstrate how preserving the full complex phase can reveal subtle but systematic timing modulations in noisy, non-stationary data. This generalizable framework offers a methodological advance for phase-coupling analysis with potential impact across neuroscience and other fields where timing dynamics are central.

## 2. METHODS

### 2.1. Phase estimation using Gabor wavelet transform

Let $x(t)$ be an fMRI time-series sampled at the repetition time (TR) $\Delta t$. The Instantaneous phase $\varphi(t)$ is estimated by convolving $x(t)$ with a complex Gabor wavelet that is both frequency and time localized:

$$g(t) = e^{(j2\pi f_0 t)} e^{\left(-\frac{t^2}{2\sigma_t^2}\right)}, \quad \sigma_t = \frac{1}{2\pi f_{bw}}, \tag{1}$$

where $f_0$ is the center frequency, and $f_{bw}$ is the half-bandwidth (Hz).

To reduce computation while retaining 99.7% of the Gaussian energy, the kernel is truncated to $t \in [-3\sigma_t, +3\sigma_t]$ [16], sampled on the TR grid ($t = k\Delta t, k \in \mathbb{Z}$). The analytic signal is

$$z(t) = (x * g)(t) = \sum_k x(k\Delta t) g(t - k\Delta t)\Delta t \tag{2}$$

with $g(t)$ normalized to unit energy to avoid edge bias. The instantaneous phase and amplitude follow as:

$$\varphi(t) = \arg[z(t)], \quad A(t) = |z(t)| \tag{3}$$

Although the wavelet itself is frequency-localized, we chose a central frequency $f_0 = 0.05Hz$ and half-bandwidth $f_{bw} = 0.02Hz$ to match the conventional $[0.03 - 0.07]Hz$ range used in Hilbert-based phase-synchrony studies, thereby isolating physiologically meaningful fMRI fluctuations while allowing direct methodological comparison [4, 17]. Relative to a Hilbert band-pass pipeline, the adaptive Gabor kernel affords sharper time–frequency localization for the non-stationary fMRI spectrum while preserving the full complex phase necessary for subsequent synchrony analysis [8].

### 2.2. Complex-valued phase synchrony & simulations

For two fMRI time-series $x(t)$ and $y(t)$ with instantaneous phases $\varphi_x(t)$ and $\varphi_y(t)$, we define the asymmetric complex phase synchrony signal as:

$$\Delta\varphi_{x\to y}(t) = \cos\{\varphi_x(t) - \varphi_y(t)\} + j\sin\{\varphi_x(t) - \varphi_y(t)\}, \ \Delta\varphi_{y\to x}(t) = \cos\{\varphi_y(t) - \varphi_x(t)\} + j\sin\{\varphi_y(t) - \varphi_x(t)\} \tag{4}$$

so that each edge retains both magnitude and signed lead–lag information.

To demonstrate the necessity of preserving the full complex angle, we generated two synthetic datasets of phase-coupled sinusoids. Each set comprised $N = 100$ signal pairs, created as $0.03\ Hz$ sine waves sampled at $1.5Hz$ for $666\ seconds$ (1000 samples). The second signal of every pair was phase-shifted by either $+60°$ (Group A) or $-60°$ (Group B); Gaussian jitter ($\sigma = 10°$) was added on each trial to mimic variability. Instantaneous phases were extracted with the Hilbert transform, yielding a stationary pair-wise phase difference $\Delta\varphi$. For every pair, we computed (i) the classical cosine synchrony $\langle\cos\Delta\varphi\rangle$ and (ii) the direction-preserving mean vector $arg\langle e^{j\Delta\varphi}\rangle$. Group differences were assessed with a two-sample $t$-test for the cosine values and Watson's $U^2$ circular-statistics test for the mean-vector angles.

Secondly, we synthesized 1000 samples of non-stationary signals to assess phase estimation accuracy under realistic fMRI conditions. A random frequency-modulated carrier was generated by integrating an instantaneous frequency that slowly "wobbled" between $0.035\ Hz$ and $0.065\ Hz$ (sinusoidal modulation sampling at $0.01\ Hz$). The carrier was multiplied by a low-frequency amplitude envelope ($0.004, 0.012, 0.022\ Hz$ components) and normalized. This frequency carrier is further encoded into a wave signal having a non-stationary phase corresponding to the simulated frequency carrier signal. Zero-mean Gaussian noise was added with $10\ dB\ SNR$ after band-passing ($0.03 - 0.07\ Hz, 0.018 - 0.098\ Hz$ transition) to match fMRI spectral content, yielding a 1000-sample (600 seconds) noisy trace. The instantaneous phase was then extracted via (i) the conventional Hilbert transform applied to the band-passed signal and (ii) the proposed complex-Gabor wavelet applied directly to the raw trace ($f_0 = 0.05\ Hz$, $f_{bw} = 0.02\ Hz$). Phase-tracking accuracy was quantified as the root mean square error between each estimate and the known ground-truth phase trajectory.

### 2.3. fMRI data & processing

We analyzed resting-state fMRI from the Function Biomedical Informatics Research Network (fBIRN) consortium. Data were acquired with a repetition time of 2 s and comprised 160 healthy controls (37.0 ± 10.9 yr, 45 F/115 M) and 151 individuals with schizophrenia (38.8 ± 11.6 yr, 36 F/115 M). Volumes underwent standard preprocessing. slice-timing correction, rigid-body realignment, MNI spatial normalization, and 6 mm FWHM Gaussian smoothing [18]. Spatially independent components were then extracted with the NeuroMark independent component analysis (ICA) pipeline [19], yielding 53 intrinsic connectivity networks (ICNs) common to all participants. The resulting ICN time courses were z-scored to unit mean and variance and subsequently submitted to the complex Gabor-wavelet phase-synchrony analysis. This study was conducted retrospectively using data collected from human participants in compliance with all relevant ethical standards.

### 2.4. CVPS states estimation

Whole-brain CVPS vectors were partitioned with standard $k$-means because of its widespread use in time-resolved fMRI analysis [11]. K-means allows us to identify recurring "states" of the CVPS metric. Each complex edge value $\Delta\varphi_{x\to y} = a + jb$ was "flattened" into two real features $[a, b]$, giving a $2E$-dimensional real vector per TR ($E =$ number of directed edges). Because $\Delta\varphi_{x\to y}$ is simply the sign-reversed complement of $\Delta\varphi_{y\to x}$, we retained one direction per pair to halve dimensionality; the omitted orientation can be recovered by sign inversion after clustering. This real-valued representation permits standard $k$-means without bespoke complex metrics while giving equal weight to cosine and sine components. Once the algorithm converged, each centroid was "unflattened" (i.e., we recombined $a$ and $b$ as $a + jb$), producing a full complex-phase template for every recurring brain state. We implemented the $k$-means algorithm with cluster numbers ranging from 1 to 10, setting the maximum number of iterations to 10,000 and using 20 random initializations to ensure robust convergence.

When each phase is embedded on the unit circle as $v(\varphi) = [\cos\varphi, \sin\varphi]$, the Euclidean distance between two angles $\varphi_1$ and $\varphi_2$ is:

$$\begin{gathered} d_E(\varphi_1, \varphi_2) \\ = \sqrt{(\cos\varphi_1 - \cos\varphi_2)^2 + (\sin\varphi_1 - \sin\varphi_2)^2} \\ d_E^2 = 2 - 2\cos(\varphi_1 - \varphi_2) = 4\sin^2\left(\frac{\Delta\varphi}{2}\right), \\ \Delta\varphi = |wrap_\pi(\varphi_1 - \varphi_2)| \in [0, \pi] \end{gathered} \tag{5}$$

Because $4\sin^2\left(\frac{\Delta\varphi}{2}\right)$ rises strictly and smoothly from 0 to 4 as $\Delta\varphi$ runs from 0 to $\pi$, minimizing the Euclidean distance, $d_E$, is equivalent to minimizing the true circular distance $\Delta\varphi$; ordinary Euclidean k-means therefore performs bona-fide angle clustering, and the arithmetic mean of the embedded points returns the circular mean of the cluster. We therefore use the Euclidean distance for the clustering with an optimal cluster number of 3, determined from the elbow criterion.

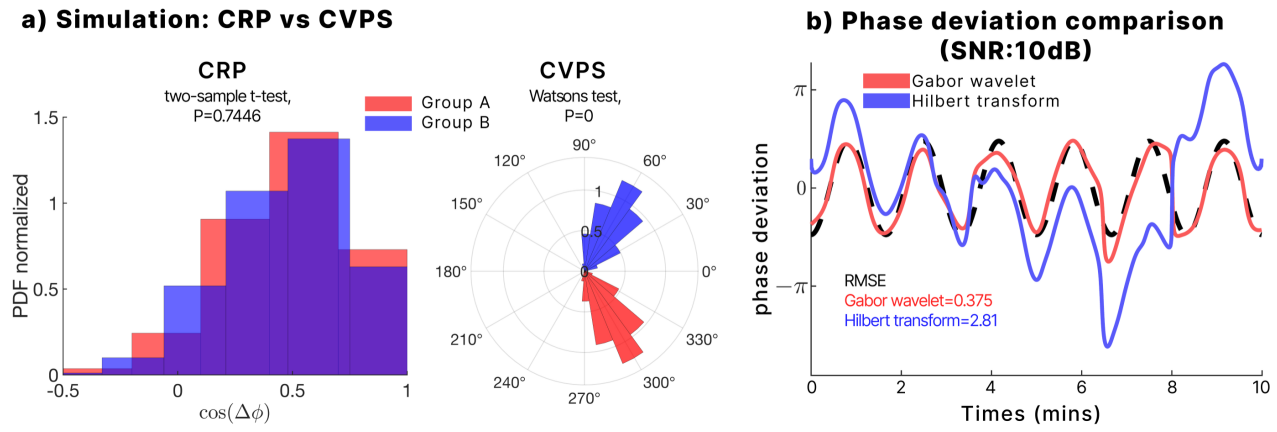

*Figure 1. (a) Validation of direction-preserving synchrony. Two synthetic groups with fixed ±60° offsets: CVPS detects separation (Watson's U², p < 0.001), whereas the cosine-only, cosine of relative phase (CRP), fails (two-sample t, p = 0.74), illustrating that CRP can mask true time-related coupling. (b) Phase-tracking accuracy for a non-stationary 10 dB SNR signal. The Gabor wavelet yields markedly lower RMSE than the Hilbert transform, indicating improved phase estimation under realistic fMRI conditions.*

### 2.5. State metrics and clinical association

From the k-means clusters, temporal behavior was summarized by three standard metrics: mean dwell time (MDT), i.e., average persistence per state; fractional rate (FR), the proportion of scan time in each state; and transition probabilities, the likelihood of switching between states [20]. In our case study, we tested whether MDT, FR, and transition probabilities were associated with chlorpromazine-equivalent dose using generalized linear models that controlled for age, sex, site, and motion, with multiple comparisons corrected by the Benjamini–Hochberg FDR. Symptom scores were added as nuisance regressors to ensure that any detected medication effects were not confounded by illness severity.

### 2.6. Computation of driver scores

To validate that the CVPS–dose association reflects a meaningful network modulation rather than a statistical artifact, we further characterize the CPZ-linked state by deriving driver scores that reveal its underlying relative-time directional backbone. Let $Z \in \mathbb{C}^{S\times E}$ be the matrix of complex-phase centroids returned by $k$-means, with $S$ states and $E$ directed edges. Each entry $Z_{s,e} = R_{s,e} \times e^{j\Delta\varphi_{s,e}}$ contains the resultant magnitude $R_{s,e} = |Z_{s,e}|$ and the phase angle $\Delta\varphi_{s,e} = argZ_{s,e}$. For a target state $s$, we quantify how well edge $e$ distinguishes that state from all other states $s'$. We first compute the smallest circular separation

$$\Delta_e = \min_{s'\neq s} \left| arg\left(e^{j\left(\Delta\varphi_{s,e}-\Delta\varphi_{s',e}\right)}\right)\right| \tag{7}$$

i.e., the minimum of the absolute angular distance between the phase in the target state and each non-target state. Using the minimum forces the score to reflect the hardest discrimination problem; an edge is only considered distinctive if it stays far from every other state, not just on average. Next, we weight this distance by a reliability term

$$S_e = \Delta_e \times min\left(R_{s,e}, \bar{R}_{\neg s,e}\right), \;\; \bar{R}_{\neg s,e} = \frac{1}{S-1}\sum_{s'\neq s} R_{s',e} \tag{8}$$

Taking the minimum of the target magnitude and the mean non-target magnitude penalizes edges that are strong in one set of states but weak in the other; only edges that are coherently expressed in both the target and comparison states receive a high weight, ensuring robustness towards noisy edges.

Edges are ranked by $S_e$ and the top 1% form the phase-critical backbone $E_s^*$. For each backbone edge $(i \to j)$ we assign a direction

$$\kappa_e = sign\left[sin\left(\Delta\varphi_{s,e}\right)\right] \in \{+1, -1\} \tag{9}$$

where $\kappa_e = +1$ means node $i$ leads node $j$. The node-level driver score is then given as

$$D_i = \sum_{e=(i\to j)\in E_s^*} \kappa_e S_e \tag{10}$$

Because only one orientation per pair is retained ($\Delta\varphi_{x\to y} = -\Delta\varphi_{y\to x}$), no opposing term appears. A positive $D_i$ value designates the region $i$ as a *leader* (net phase lead), while a negative value designates it as a *follower* (net lag) within the state $s$. This circular-distance-weighted, orientation-aware aggregation thus provides a novel driver mapping of each recurring complex-phase state.

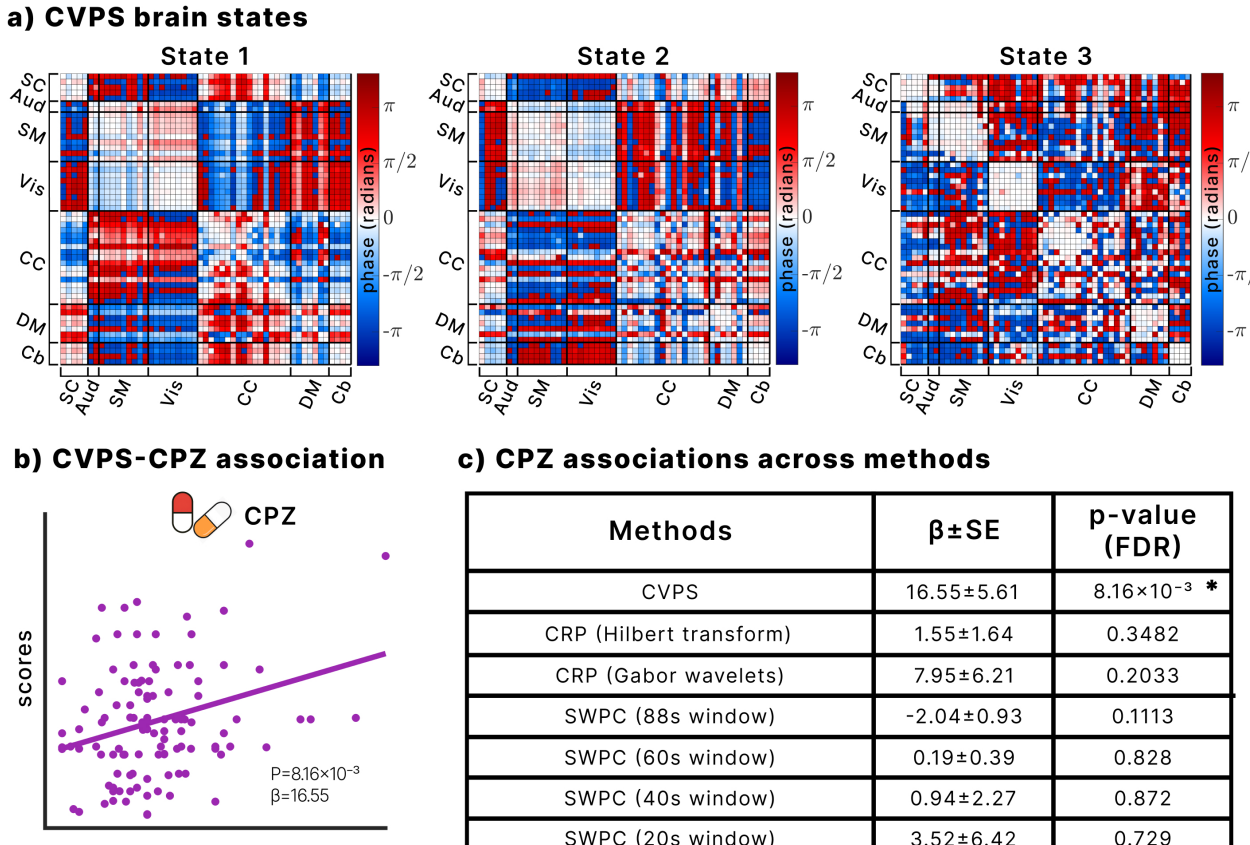

| Methods | β±SE | p-value (FDR) |
|---|---|---|
| CVPS | 16.55±5.61 | $8.16\times10^{-3}$ * |
| CRP (Hilbert transform) | 1.55±1.64 | 0.3482 |
| CRP (Gabor wavelets) | 7.95±6.21 | 0.2033 |
| SWPC (88s window) | -2.04±0.93 | 0.1113 |
| SWPC (60s window) | 0.19±0.39 | 0.828 |
| SWPC (40s window) | 0.94±2.27 | 0.872 |
| SWPC (20s window) | 3.52±6.42 | 0.729 |

*Figure 2. (a) Centroids of the three CVPS states (angles in radians). (b) In schizophrenia, mean dwell time (MDT) in State 2 increases with CPZ dose after covariate adjustment and FDR correction. (c) Cosine-only PS and sliding-window Pearson correlation (SWPC) show no significant associations, highlighting that only CVPS tracks medication effects in a single cross-sectional scan.*

## 3. RESULTS

### 3.1. Simulation benchmark preserving relative time directionality and robust phase estimation

Figure 1 illustrates the advantages of CVPS. In the first test, two synthetic groups with ±60°, phase shifts were generated. CVPS correctly separates them (Watson $U^2, p << 0.001$), whereas the cosine-only measure fails ($p = 0.74$; Fig. 1a). In the second test with a frequency-modulated signal embedded in 10 $dB$ Gaussian noise, Gabor-wavelet phase estimation tracked the true trajectory far more accurately than the Hilbert transform (RMSE 0.38 vs 2.81 across 1000 runs; Fig. 1b). These results highlight the superiority of our proposed pipeline, showing that it provides more faithful phase estimation under nonstationary and noisy conditions such as brain fMRI signals, while preserving both cosine and sine components so that direction of timing coupling is retained rather than collapsed into a symmetric coupling strength measure.

### 3.2. CPZ association of CVPS vs conventional methods

Figure 2 summarizes the complex-phase brain states and their relation to antipsychotic load. Fig. 2 (a) shows the three CVPS centroids expressed in radians. In Fig. 2 (b), MDT in state 2 increases with chlorpromazine-equivalent dose ($\beta = 16.55, p = 8.16 \times 10^{-3}$) after adjustment for age, sex, motion, site, and symptom scores, indicating that the effect reflects medication rather than symptom severity; we therefore refer to State 2 as the "CPZ state."

To evaluate whether this effect could be explained by more conventional pipelines, we compared CVPS against cosine-only phase synchronization using both Hilbert- and Gabor-based phase extraction, as well as against sliding-window Pearson correlation (SWPC), the most widely used dFC measure. For SWPC, we followed prior work by testing multiple window lengths (88 s to match the −3 dB point of the 0.01 Hz low-band cutoff [21], as well as 60 s, 40 s, and 20 s) to ensure fairness of comparison. All methods were clustered with k-means using Euclidean distance, fixing k = 3 to match the optimal solution from the CVPS pipeline, ensuring fair comparison across methods. Similar generalized linear models with identical covariates were applied to test for CPZ associations.

Fig. 2 (c) shows the best associations recovered from each alternative method. Neither cosine-only PS nor SWPC yielded significant effects of CPZ dose across any state or window size. This contrast demonstrates that the association is not due simply to the use of the Gabor wavelet, but to the joint benefit of (i) more accurate phase estimation and (ii) retaining both cosine and sine components of the phase difference. This highlights that only CVPS is sensitive enough to track medication modulation in a single cross-sectional scan, whereas conventional approaches fail to detect this timing-dependent effect.

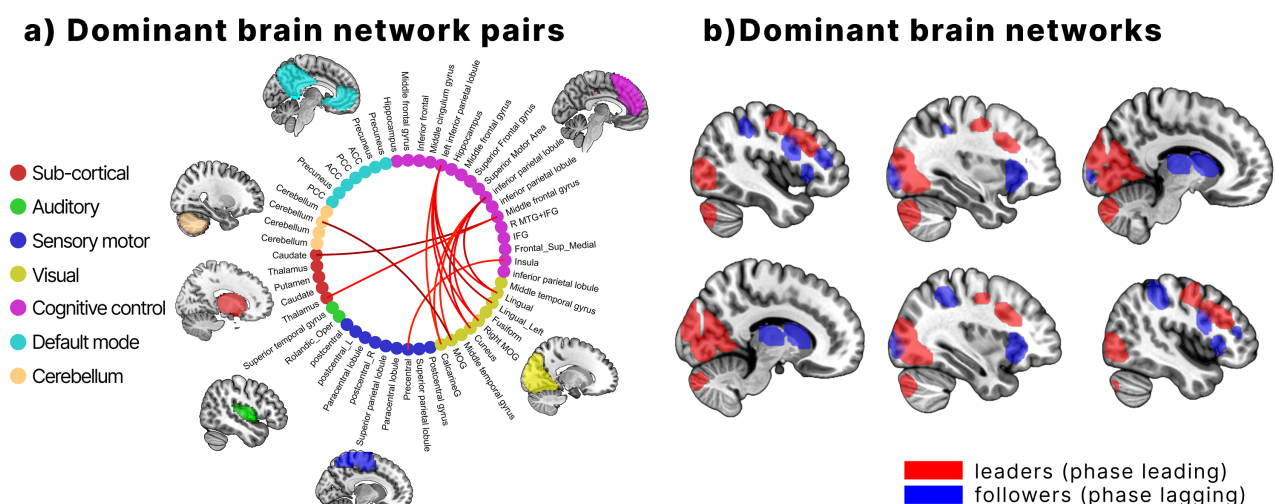

*Figure 3 (a) Top 1 % of phase-discriminative edges that characterize the CPZ-associated State 2, summarized across seven canonical brain domains. (b) Node-level driver maps for State 2, with regions that lead the phase flow (leaders) shown in red and regions that lag (followers) shown in blue.*

### 3.3. CVPS validation through pharmacological circuit mechanisms

Figure 3 characterizes the network architecture of the CPZ-linked state using our circular-distance-weighted driver score analysis. In Fig. 3 (a), the top-ranked 1% of discriminative edges are shown, revealing a highly consistent pattern. The strongest edges converge on the left inferior parietal lobule (IPL) and adjacent parietal regions, each connecting to early visual cortices, including calcarine, cuneus, lingual, and middle occipital gyri. Additional high-scoring connections include the middle frontal gyrus projecting to the caudate and the middle occipital gyrus projecting to the cerebellum.

Fig. 3 (b) translates these edge-level findings into directional flow maps of leaders and followers. Visual regions consistently lead their parietal partners, suggesting that sensory evidence is time-stamped in early visual cortex before being integrated by association areas. This observation mirrors prior work showing that visual network signals "shrink" relative to associative hubs [22, 23] and aligns with resting-state MEG and spectral dynamic causal modeling studies demonstrating robust occipital-to-parietal information flow [24-26]. Such a bottom-up stream is consistent with the idea that antipsychotics dampen parietal over-responsiveness and stabilize sensory evidence before higher-order appraisal [27]. The fact that CVPS recovers this expected bottom-up pattern validates that the CPZ state reflects genuine circuit modulation.

In parallel, the middle frontal gyrus leads both the caudate and thalamus, consistent with dopaminergic $D_2$-receptor blockade reducing striatal drive and shifting temporal control toward the cortex [15], further validating that CVPS recovers pharmacologically plausible modulation. The cerebellar loop inversion, where the cerebellum leads the middle occipital gyrus, echoes pharmacological fMRI findings implicating cerebellar modulation of sensory gating [28]. These matches to known pharmacological effects confirm that CVPS does not simply discover arbitrary clusters but recovers biologically plausible leader–follower hierarchies shaped by antipsychotic action.

While these lead–lag relations align with known pharmacological targets, it is important to note that fMRI phase offsets reflect a combination of neural timing and hemodynamic latencies, not purely neural conduction. Nevertheless, prior work shows that such latencies are reproducible and behaviorally meaningful [29, 30], supporting their interpretation as pharmacodynamically relevant timing effects.

Together, these relative timing patterns provide mechanistic insight into the CPZ state: bottom-up occipital-to-parietal flow stabilizing sensory integration, paired with top-down prefrontal-to-subcortical leadership. This dual organization provides a strong validation that preserving the full complex representation of phase synchrony captures meaningful, pharmacodynamically relevant circuit retiming.

## 4. CONCLUSION

This work introduces complex-valued phase synchrony with a Gabor-wavelet phase extractor as a robust alternative to conventional Hilbert/cosine pipelines, preserving both magnitude and signed lag for direction-sensitive coupling. Applied to resting-state fMRI, CVPS uncovers a reproducible state whose persistence scales with antipsychotic dose in a single cross-sectional scan, a result not captured by cosine-only phase synchrony or dynamic functional connectivity across several windows. These findings establish CVPS as a generalizable framework for detecting pharmacologically relevant timing modulations and, more broadly, for quantifying relative time directional coupling in complex, noisy, and non-stationary systems.

## 5. ACKNOWLEDGEMENT

This work was supported by the National Institutes of Health (NIH) grant (R01MH123610) and the National Science Foundation (NSF) grant #2112455.

## 6. REFERENCES

1. Granger, C.W., *Investigating causal relations by econometric models and cross-spectral methods.* Econometrica: journal of the Econometric Society, 1969: p. 424-438.
2. Allen, E.A., et al., *Tracking whole-brain connectivity dynamics in the resting state.* Cereb Cortex, 2014. **24**(3): p. 663-76.

3. Mohanty, R., et al., *Rethinking Measures of Functional Connectivity via Feature Extraction.* Scientific Reports, 2020. **10**(1): p. 1298.
4. Glerean, E., et al., *Functional magnetic resonance imaging phase synchronization as a measure of dynamic functional connectivity.* Brain Connect, 2012. **2**(2): p. 91-101.
5. Honari, H., et al., *Evaluating phase synchronization methods in fMRI: A comparison study and new approaches.* NeuroImage, 2020. **228**: p. 117704 - 117704.
6. Wiafe, S.-L., et al., *Studying time-resolved functional connectivity via communication theory: on the complementary nature of phase synchronization and sliding window Pearson correlation.* bioRxiv, 2024: p. 2024.06. 12.598720.
7. Pedersen, M., et al., *On the relationship between instantaneous phase synchrony and correlation-based sliding windows for time-resolved fMRI connectivity analysis.* NeuroImage, 2018. **181**: p. 85-94.
8. Li, D., et al., *Phase synchronization with harmonic wavelet transform with application to neuronal populations.* Neurocomputing, 2011. **74**(17): p. 3389-3403.
9. Lottman, K.K., et al., *Risperidone Effects on Brain Dynamic Connectivity-A Prospective Resting-State fMRI Study in Schizophrenia.* Front Psychiatry, 2017. **8**: p. 14.
10. Salman, M.S., et al., *Decreased cross-domain mutual information in schizophrenia from dynamic connectivity states.* Frontiers in Neuroscience, 2019. **13**: p. 457620.
11. Damaraju, E., et al., *Dynamic functional connectivity analysis reveals transient states of dysconnectivity in schizophrenia.* NeuroImage: Clinical, 2014. **5**: p. 298-308.
12. Sambataro, F., et al., *Treatment with olanzapine is associated with modulation of the default mode network in patients with Schizophrenia.* Neuropsychopharmacology, 2010. **35**(4): p. 904-12.
13. Ishii, T., et al., *Anatomical relationships between serotonin 5-HT2A and dopamine D2 receptors in living human brain.* PLoS One, 2017. **12**(12): p. e0189318.
14. Sommer, I.E.C., et al., *Successful treatment of intractable visual hallucinations with 5-HT ((2A)) antagonist ketanserin.* BMJ Case Rep, 2018. **2018**.
15. Peters, S.K., K. Dunlop, and J. Downar, *Cortico-Striatal-Thalamic Loop Circuits of the Salience Network: A Central Pathway in Psychiatric Disease and Treatment.* Front Syst Neurosci, 2016. **10**: p. 104.
16. Oppenheim, A.V., *Discrete-time signal processing*. 1999: Pearson Education India.
17. Achard, S., et al., *A resilient, low-frequency, small-world human brain functional network with highly connected association cortical hubs.* Journal of Neuroscience, 2006. **26**(1): p. 63-72.
18. Penny, W.D., et al., *Statistical parametric mapping: the analysis of functional brain images*. 2011: Elsevier.
19. Du, Y., et al., *NeuroMark: An automated and adaptive ICA based pipeline to identify reproducible fMRI markers of brain disorders.* Neuroimage Clin, 2020. **28**: p. 102375.
20. Iraji, A., et al., *Tools of the trade: estimating time-varying connectivity patterns from fMRI data.* Social Cognitive and Affective Neuroscience, 2020. **16**(8): p. 849-874.
21. Faghiri, A., et al., *A unified approach for characterizing static/dynamic connectivity frequency profiles using filter banks.* Network Neuroscience, 2021. **5**(1): p. 56-82.
22. Wiafe, S.-L., et al., *Capturing Stretching and Shrinking of Inter-Network Temporal Coupling in FMRI Via WARP Elasticity*. 2024. 1-4.
23. Wiafe, S.-L., et al., *The dynamics of dynamic time warping in fMRI data: a method to capture inter-network stretching and shrinking via warp elasticity.* Imaging Neuroscience, 2024.
24. Hillebrand, A., et al., *Direction of information flow in large-scale resting-state networks is frequency-dependent.* Proceedings of the National Academy of Sciences, 2016. **113**(14): p. 3867-3872.
25. Razi, A., et al., *Construct validation of a DCM for resting state fMRI.* Neuroimage, 2015. **106**: p. 1-14.
26. Tewarie, P., et al., *Integrating cross-frequency and within band functional networks in resting-state MEG: A multi-layer network approach.* Neuroimage, 2016. **142**: p. 324-336.
27. Keedy, S.K., et al., *An fMRI study of visual attention and sensorimotor function before and after antipsychotic treatment in first-episode schizophrenia.* Psychiatry Res, 2009. **172**(1): p. 16-23.
28. Minassian, A., D. Feifel, and W. Perry, *The relationship between sensorimotor gating and clinical improvement in acutely ill schizophrenia patients.* Schizophrenia Research, 2007. **89**(1): p. 225-231.
29. Mitra, A., et al., *Lag structure in resting-state fMRI.* J Neurophysiol, 2014. **111**(11): p. 2374-91.
30. Rangaprakash, D., R.L. Barry, and G. Deshpande, *The confound of hemodynamic response function variability in human resting-state functional MRI studies.* Front Neurosci, 2023. **17**: p. 934138.